\documentclass[twocolumn,floatfix,superscriptaddress,a4paper,showpacs,showkeys,reprint,prc,nofootinbib]{revtex4-2}
\usepackage{epsfig}
\usepackage{longtable}
\usepackage{latexsym}
\usepackage[colorlinks=true,linktocpage=true,linkcolor=blue,citecolor=blue,allcolors=blue]{hyperref}
\usepackage{url}
\usepackage[utf8]{inputenc}
\usepackage{enumerate}
\usepackage{color}
\usepackage{xcolor}
\usepackage{setspace}
\usepackage{amsmath}
\usepackage{amssymb}
\usepackage[english]{babel}
\usepackage{url}

\begin{document}

\title{Ultra fast, event-by-event heavy-ion simulations for next generation experiments}

\author{Manjunath Omana Kuttan}
\email{manjunath@fias.uni-frankfurt.de}
\affiliation{Frankfurt Institute for Advanced Studies, Ruth-Moufang-Str. 1, D-60438 Frankfurt am Main, Germany}
\affiliation{Xidian-FIAS International Joint Research Center, Giersch Science Center, D-60438 Frankfurt am Main, Germany}

\author{Kai Zhou}
\email{zhoukai@cuhk.edu.cn}
\affiliation{School of Science and Engineering, The Chinese University of Hong Kong, Shenzhen, P.R. China}
\affiliation{Frankfurt Institute for Advanced Studies, Ruth-Moufang-Str. 1, D-60438 Frankfurt am Main, Germany}

\author{Jan Steinheimer}
\affiliation{GSI Helmholtzzentrum f\"ur Schwerionenforschung GmbH, Planckstr. 1, D-64291 Darmstadt, Germany}

\affiliation{Frankfurt Institute for Advanced Studies, Ruth-Moufang-Str. 1, D-60438 Frankfurt am Main, Germany}

\author{Horst Stoecker}
\affiliation{Frankfurt Institute for Advanced Studies, Ruth-Moufang-Str. 1, D-60438 Frankfurt am Main, Germany}
\affiliation{GSI Helmholtzzentrum f\"ur Schwerionenforschung GmbH, Planckstr. 1, D-64291 Darmstadt, Germany}
\affiliation{Institut f\"{u}r Theoretische Physik, Goethe Universit\"{a}t Frankfurt, Max-von-Laue-Str. 1, D-60438 Frankfurt am Main, Germany}

\date{\today}
\begin{abstract}

We present a novel deep generative framework that uses probabilistic diffusion models for ultra fast, event-by-event simulations of heavy-ion collision output. This new framework is trained on UrQMD cascade data to generate a full collision event output containing 26 distinct hadron species. The output is represented as a point cloud, where each point is defined by a particle's momentum vector and its corresponding species information (ID). Our architecture integrates a normalizing flow-based condition generator that encodes global event features into a latent vector, and a diffusion model that synthesizes a point cloud of particles based on this condition. A detailed description of the model and an in-depth analysis of its performance is provided. The conditional point cloud diffusion model learns to generate realistic output particles of collision events which successfully reproduce the UrQMD distributions for multiplicity, momentum and rapidity of each hadron type. The flexible point cloud representation of the event output preserves full event-level granularity, enabling direct application to inverse problems and parameter estimation tasks while also making it easily adaptable for accelerating any event-by-event model calculation or detector simulation.\end{abstract}

\maketitle
\section{Introduction}
The study of strongly interacting matter under various thermodynamic conditions, governed by quantum chromodynamics (QCD), lies at the heart of relativistic heavy-ion collision experiments worldwide \cite{Stoecker:1986ci,Fukushima:2010bq}. The temperature and density reached in these collisions are varied by tuning the collision energy to explore the largely conjectured QCD phase diagram. While very high collision energies ($\sqrt{s_{\mathrm{NN}}} \gtrapprox $ 200~GeV) are expected to produce systems with vanishing net-baryon densities, high density QCD matter ($\gtrapprox$ 2 times nuclear saturation density $n_0$) can be explored at intermediate collision energies   ($\sqrt{s_{\mathrm{NN}}} \approx $ 2-10~GeV). At high baryon densities, the phase structure of QCD is speculated to contain several interesting structures such as a potential first order-phase transition from hadron to parton matter and a critical end point.

Lattice QCD calculations \cite{Aoki:2006we, Borsanyi:2010cj, HotQCD:2014kol,Cheng:2007jq,Bazavov:2011nk,PACS-CS:2008bkb} have shown that, at vanishing net-baryon densities, hadronic matter transforms to a deconfined quark-gluon phase via a smooth crossover. However, lattice QCD calculations at finite baryon densities remain intractable due to the fermionic sign problem, necessitating reliance on effective models of QCD and advanced computational tools that are closely connected to the experiments to explore high baryon density QCD matter.

The experimental programmes of STAR-FXT and STAR-BES at RHIC, CBM at SIS-100, HADES at SIS-18, CEE at HIAF and MPD at NICA are dedicated to studying intermediate energy nuclear collisions  \cite{HADES:2009aat,Yang:2017llt, STAR:2010vob, STAR:2017sal,Meehan:2016qon,CBM:2016kpk,Yang:2013yeb,Golovatyuk:2016zps}. With unprecedented event rates and measurement precision, these experiments provide a unique opportunity to investigate the mostly unknown, high density region of QCD phase diagram using rare observables and novel methods. The experimental efforts are complemented by various approaches for a reliable theoretical description of relativistic, moderate energy heavy-ion collisions.

A well motivated, microscopic, and non-equilibrium description of heavy ion collisions at these intermediate-energies is provided by well tested transport models \cite{Bass:1998ca,Bleicher:1999xi,Cassing:2009vt,Nara:1999dz}. An alternative strategy is to adopt a hybrid framework, in which the hot and dense hadron matter phase is evolved via hydrodynamics while the initial and final stages, including event-by-event fluctuations, are treated through transport theory \cite{Petersen:2008dd}. A key advantage of such a hybrid description is the explicit inclusion of various, unknown types of QCD transitions or critical endpoints through different Equations of State (EoS) modeled to drive the hydrodynamic evolution. However, important non-equilibrium phenomena such as the influence of the EoS in the initial compression phase are neglected in such an approach. Recently, it was shown that a self-consistent transport description of the entire evolution with any assumed EoS can be achieved by incorporating a density dependent potential within the QMD part of ultra-relativistic quantum molecule dynamics (UrQMD) \cite{OmanaKuttan:2022the,Steinheimer:2023xzs}. Building on this concept, both momentum and particle dependence for the EoS can also be realized within the UrQMD model as demonstrated in \cite{Steinheimer:2024eha}. Similar developments are also attempted in other transport model approaches \cite{Oliinychenko:2022uvy,Mohs:2024gyc,Nara:2021fuu}.

The experimental determination of the EoS remains extremely challenging. Several observables \cite{Stoecker:1986ci,Hofmann:1976dy,Stoecker:2004qu,Stephanov:1998dy,Hatta:2003wn,osti_4061525,hofmann1975report} have been conjectured to be sensitive to the EoS of the hot, dense matter created in heavy-ion collisions. However, an unambiguous determination of the EoS has not been presented yet. A systematic comparison of extensive model calculations with needed experimental data for multiple observables is necessary to infer the EoS  which provides the best description for experimental data. Such analyses are typically conducted using a computationally intensive Bayesian inference procedure. However, current models for event-by-event simulations are not efficient enough to calculate different observables with sufficient statistical significance
for several thousands of different parameter sets which are necessary to construct a reliable Bayesian posterior. As a result, current Bayesian inference methods rely on machine learning based surrogate models which serve as fast emulators, mapping the model parameters to various observables \cite{Bernhard:2016tnd,JETSCAPE:2020mzn,Pratt:2015zsa,Nijs:2020roc,OmanaKuttan:2022aml,Oliinychenko:2022uvy,Huth:2021bsp,Cheng:2023ucp}. This strategy is effective for a small number of computationally inexpensive observables such as integrated flow coefficients. However, the number of simulations necessary to generate sufficient training data quickly makes this approach infeasible if the analysis were to include a large number of observables requiring large number of events to compute, e.g. higher order fluctuations or multi-differential flow of various hadronic species. Hence, further advancements in both the computational efficiency of simulation models and in the development of surrogate models are crucial for accurate extraction the EoS from intermediate-energy heavy-ion collision data.

The present work therefore  primarily addresses the issue with a Deep Learning (DL) based solution for accelerating the event-by-event simulation of heavy-ion collisions. Deep Learning techniques have been widely used for various physics studies of relativistic heavy-ion collisions \cite{Zhou:2023pti,OmanaKuttan:2020brq, OmanaKuttan:2020btb, OmanaKuttan:2023bnb, Pang:2016vdc, Du:2019civ,Du:2020pmp,He:2023zin,Li:2020qqn,Mallick:2022alr,Song:2021rmm,Hirvonen:2023lqy,Sergeev:2020fir,Jiang:2021gsw,Murali:2024xnf,Aarts:2025gyp}. Recently, generative modeling, an unsupervised deep learning technique that learns to generate realistic synthetic samples based on patterns learned from the training data, has gained considerable attention \cite{Zhou:2018ill,Cranmer:2023xbe}. Generative models have served successfully in several tasks like natural language processing and image synthesis.

Currently, generative methods are also being explored  in heavy-ion physics to generate various final state spectra \cite{Sun:2024lgo, Huang:2018fzn} and the calorimeter response in experiment \cite{Torbunov:2024iki}. While these methods are useful in specific scenarios, our goal is to develop a model that generates a complete, event-by-event, heavy-ion collision output. Such a framework can be used for rapid computation of any observable as predicted by a given theoretical model, thus eliminating the need to train separate surrogate models for each observable. This advantage is particularly valuable in comprehensive Bayesian analyses, where numerous, computationally intensive observables (e.g., higher-order fluctuations or multi-differential flow) must often be considered for constraining the EoS.
Moreover, such a generative framework can enable real-time model-data comparisons during experiments, and can be combined with optimization techniques such as gradient descent to identify model inputs which best describe the data,offering fast and robust alternatives for exploiting large amounts of experimental data and extracting various properties of the system. 

This work lays the groundwork for an ultra fast AI-driven event-by event simulation model for relativistic heavy-ion collisions. As a first step, we illustrate how this goal can be achieved using training data simulated from the UrQMD cascade model for a fixed beam energy and centrality. Complementing the main results of our method presented in \cite{OmanaKuttan:2024mwr}, a detailed account of the model structure, its implementation and a comprehensive analysis of its performance is presented here.

\section{The UrQMD model and point cloud representation}
UrQMD provides a well established non-equilibrium transport description of relativistic heavy-ion collisions over a wide range of collision energies, from SIS to RHIC and LHC. Within UrQMD, hadrons are propagated according to Hamilton's equations of motion and interact via stochastic binary scatterings, color string formation and resonance excitation and decays. The default setup of UrQMD is a hadronic cascade model which has the effective EoS of a Hadron Resonance Gas (HRG) \cite{Bravina:2008ra}. There is an option to turn on nuclear potentials and an intermediate hydrodynamic stage to include arbitrary EoSs in the evolution of the system. However, for the present demonstrative study, we restrict ourselves to the cascade mode of UrQMD.

In UrQMD, both projectile and target nuclei are boosted towards each other with impact parameter $b$ less than the sum of the projectile and target radii i.e., $b<R_{p}+R_{t}$. The initial positions of nucleons in each nucleus follows a Woods-Saxon  radial distribution. The nucleons are assigned random Fermi momenta. A geometric interpretation of the cross section is employed to describe the interactions of hadrons traveling in straight trajectories. The UrQMD cascade describes successfully the final state hadron spectra over a wide range of energies and the projectile and target masses.

The UrQMD cascade propagates all the nucleons and produces hadrons until all elastic and inelastic interactions cease. Then it outputs the complete list of all final state particles including their momentum information. An event output from UrQMD can therefore be considered as a point cloud in phase space of final state particles. A point cloud is a collection of points in some feature space with no specific order. Hence, an Artificial Intelligence (AI) model that aims to generate "UrQMD-like" events must deliver precise and complete, particle-by-particle information in such point cloud format rather than aggregate information in the form of histograms or spectra.

\section{HEIDi: Heavy-ion Events Through Intelligent Diffusion}
Generative modeling of point clouds is an active area of research in AI, motivated by their extensive use in fields where data naturally takes the form of unordered point sets. Electronically collected data often have an inherent point cloud structure. 
Point clouds are the natural form for representing LIDAR data, which are used widely in 3D terrain mapping and autonomous driving, for 3D medical scans in medical imaging and for 
processing any sensor-based data in general. Generative models for point clouds find applications in point cloud completion, synthesis, de-noising, and segmentation, altering drastically how electronic data is processed across various research fields. From generative adversarial networks to autoregressive point cloud generation, point cloud diffusion and flow based models, different approaches for generating point clouds have been proposed \cite{pmlr-v80-achlioptas18a,Yang_2019_ICCV,guo2020deep,Luo_2021_CVPR, sun2020pointgrow}. 

Diffusion models, in particular, starts by sampling from a simple well known distribution and then iteratively denoise them to generate synthetic samples from the original data distribution. Diffusion models have been explored in particle physics for generating calorimeter responses \cite{Mikuni:2022xry, Amram:2023onf,Acosta:2023zik,Mikuni:2023tqg,Buhmann:2023bwk}, jets \cite{Butter:2023fov,Leigh:2023zle,Buhmann:2023zgc,Buhmann:2023pmh,Mikuni:2023dvk,Leigh:2023toe}, and electron- proton scattering events \cite{Araz:2024bom,Devlin:2023jzp} at high energy particle collider facilities worldwide.

In this work, we present a probabilistic, conditional diffusion model based on \cite{Luo_2021_CVPR}, to generate point clouds representing the complete heavy-ion collision event output, hereafter referred to as \emph{\textbf{HEIDi}: \textbf{H}eavy-ion \textbf{E}vents through \textbf{I}ntelligent \textbf{Di}ffusion}.

\emph{HEIDi} is trained here by using 18000 Au-Au collision events at $E_{lab}$= 10 $A$GeV with fixed impact parameter $b$=1 fm. In order to fix the input dimensions of the training data, each event is represented as a point cloud $\mathbf{X} = \{ \mathbf{x}_i \}_{i=1}^{1084}$ containing 1084 particles (points) where 1084 is a number larger than the maximum event multiplicity of the training dataset. Events with multiplicity less than 1084 are filled up with zeros, which denote fake particles that maintain consistent input dimensions. Each particle $\mathbf{x}_i$ in the point cloud has 32 attributes, of which 3 attributes represent the three components of the momentum vector $p_{x_i}, p_{y_i}$ and $p_{z_i}$ while the remaining 29 attributes denote the one-hot encoded particle information ($\text{ID}_i$): 
\begin{equation}
 \mathbf{x}_i = \{ \mathbf{p}_{i}, \text{ID}_i \},  \quad \mathbf{p}_{i} = (p_{x_i}, p_{y_i}, p_{z_i}).
\end{equation}

The one-hot encoding $\text{ID}_i$ creates a 29 dimensional vector denoting unique IDs for the following 26 different hadron types: $ \bar{p}, \bar{n}, K^-, \bar{K}^0, \bar{\Lambda}^0, \bar{\Sigma}^-, \bar{\Sigma}^0, \bar{\Sigma}^+, \bar{\Xi}^0, \bar{\Xi}^+, \bar{\Omega}, n, p, \pi^-, \pi^0, \pi^+, \\ \eta^0, K^0, K^+, \Lambda^0, \Sigma^-, \Sigma^0, \Sigma^+, \Xi^-, \Xi^0, \Omega^-$ as well as  IDs for spectator\footnote{Nucleons which did not undergo any elastic or inelastic interaction are defined as  spectator nucleons.} neutrons $n^\prime$, spectator protons $p^\prime$ and the fake particles $0$. In this way we create a point cloud of complete collision event output represented as a $1084 \times 32$ array for each event.

\begin{figure*}[]
   \includegraphics[width=\textwidth]{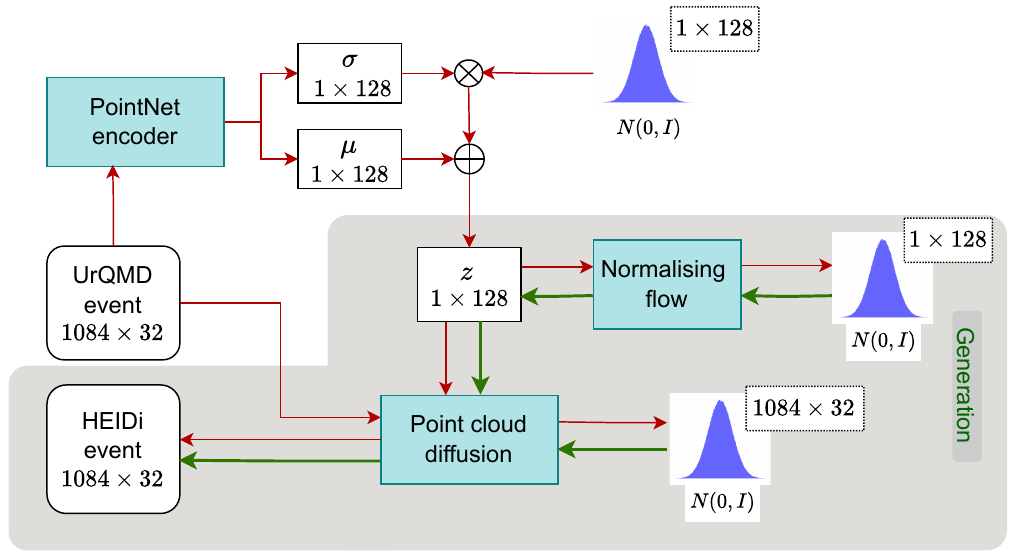}
   \caption{(Color online) \emph{HEIDi}: network structure. The red and green arrows show the flow of information during training and generation, respectively. Only the parts shaded in gray are used during the generation process. A collision event output is represented by a point cloud containing 1084 points where each point is a final state particle. $\sigma$ and $\mu$ represent the mean and standard deviation, respectively, of a 128 dimensional Gaussian distribution used to construct the latent condition vector $z$. The generation process starts with random samples from the standard normal distribution $\mathcal{N}(0,I)$  (mean= zero vector, covariance = identity matrix $I$), which is processed by the normalizing flow-based decoder and the diffusion model to generate the final state event output point cloud. The labels '$1 \times 128$' and '$1084 \times 32$' refer to the dimensionality of the standard normal distribution from which the random samples are drawn.}
   \label{heidi}
 \end{figure*}

The structure of \emph{HEIDi} is illustrated in figure \ref{heidi}. \emph{HEIDi} comprises 2 major components: a) PointNet-based  \cite{qi2017pointnet}  encoder, with a normalizing flow-based \cite{dinh2016density} decoder, for generating the condition vector $z$ that encoder various global event features and correlations, b) a diffusion model \cite{ho2020denoising} that takes the vector $z$ as condition to generate a point cloud of collision output.

Our first objective is to generate a condition $z$ that informs the diffusion model of various correlations or features that should be present in the generated data. For this, we need a way to map various correlations in an event in the training data into a latent variable $z$ and also learn the probability distribution of this latent variable. After training, by sampling from this probability distribution, we can then generate the condition $z$. This is achieved by the use of a PointNet-based encoder and a normalizing flow-based decoder.

\subsection{PointNet-based encoder}
 
PointNet is a network designed to learn efficiently from point cloud data, while respecting its permutation invariance. Applications of PointNet-based models in heavy-ion collisions have already been explored in several works \cite{OmanaKuttan:2020brq,OmanaKuttan:2020btb,OmanaKuttan:2023bnb,OmanaKuttan:2021axp,Guo:2023zfk}. PointNet constructs order invariant features of a point cloud, by extracting features of each point separately and converting them into a global feature of the entire point cloud, by using a symmetric mathematical function such as sum or average of all the point-wise features:
\begin{equation}
\mathcal{Y}=f(\mathbf{X})=g(h(\mathbf{{x_1}}),h(\mathbf{{x_2}})...h(\mathbf{x_{1084}})).
\end{equation}

Here, $\mathcal{Y}$  is an order invariant feature of the point cloud, $h$ is the point-wise feature extraction operation usually modeled as a neural network, using 1-D convolution operations or a multi layered perceptron with shared weights. $g$ is the symmetric operation applied on the extracted point-wise features.

For every UrQMD event in the training dataset, our PointNet-based encoder predicts two global feature vectors of dimensions 1 $\times$ 128 each, which represent the mean $\mu$ and the standard deviation $\sigma$ of a 128-dimensional Gaussian distribution. Random samples generated from this Gaussian $\mathcal{N}(\mu,\sigma)$ has to be used as the condition vector for the diffusion model to optimize the encoder parameters and learn the correct encoding. To achieve this, the gradients of loss/objective function $\mathcal{L}$ (a mathematical function which quantifies the difference between the expected and actual outputs ) with respect to model parameters need to be computed. Since random sampling is  non-deterministic and non-differentiable, the well-known reparameterization trick \cite{kingma2013auto} is used to ensure the flow of gradients through the network. Here, the random sampling process is isolated into a fixed distribution, typically a standard normal distribution $\mathcal{N}(0,I)$  (mean= zero vector, covariance = identity matrix $I$) and the latent encoding is parameterized as the variable $z$ given by: 
\begin{equation}
    z = \mu + \sigma \cdot \epsilon, \quad \text{where} \quad \epsilon \sim \mathcal{N}(0,I).
\end{equation}

The encoder neural network with parameters $\phi$ estimates the approximate probability distribution $ \tilde{q}_{\phi}(z | \mathbf{X}) $ for condition $z$ given the point cloud $\mathbf{X}$ of UrQMD collision output as a Gaussian with mean $ \mu_{\phi}$ and covariance $\Sigma_{\phi}$ predicted by the network: 
\begin{equation}
    i.e., \tilde{q}_{\phi}(z | \mathbf{X}) = \mathcal{N} \left(z \middle| \mu_{\phi}(\mathbf{X}), \Sigma_{\phi}(\mathbf{X}) \right).
    \label{eqphi}
\end{equation}

\subsection{Normalizing flow based decoder}
To complete the pipeline which involves generating the conditional vector $z$, we still need to know the 
probability density $p(z)$. We use a normalizing flow-based model to facilitate the sampling of $z$ from $p(z)$. Instead of directly learning the potentially complex, high dimensional $p(z)$ distribution, the normalizing flow model with parameters $\alpha$ learns bijective transformations $\mathcal{B}_{\alpha}(z)$ which convert samples from $p(z)$ as generated by the PointNet based encoder, into samples $w$ following a simple, well known distribution, such as the standard normal distribution $\mathcal{N}(0,I)$. Thus, after training, we can invert the process to sample from the prior normal distribution and transform it via $\mathcal{B}^{-1}_{\alpha}(w)$ into samples from $p(z)$. For more details on normalizing flow models, see \cite{papamakarios2021normalizing}. The bijective nature of the transformations modeled by this neural network would allow us to invert the transformations and extract the exact probability for $z$ as
\begin{equation}
p(z)=p(w) \cdot \left| \det \frac{\partial \mathcal{B}_{\alpha}^{-1}}{\partial w} \right|^{-1},
\label{eqz}
\end{equation}
where $w=\mathcal{B}_{\alpha}(z)$ and $\det \frac{\partial \mathcal{B}_{\alpha}^{-1}}{\partial w}$ is the Jacobian determinant of $\mathcal{B}_{\alpha}^{-1}$.

\subsection{Point cloud diffusion }
The final step is to generate a point cloud of collision output based on the condition $z$. For this, we employ a probabilistic diffusion model. For different implementations and applications of diffusion models, see \cite{croitoru2023diffusion,yang2023diffusion,Wang:2023sry,Wang:2023exq}. The learning process in a diffusion model comprises a forward diffusion step and a reverse anti-diffusion step.

In the forward diffusion step, a controlled Gaussian noise is gradually added to the UrQMD data over several time steps until the point cloud is transformed into pure noise. The original UrQMD generated point cloud $\mathbf{X} =\mathbf{X}^0 = \{ \mathbf{x}_i^0 \}_{i=1}^{1084}$ is considered as the initial, undisturbed state of the point cloud at time step $t=0$. At each subsequent time step, a Gaussian noise is added to the previous state of each point in the point cloud to obtain the current state of the point. This results in states $\mathbf{x}_i^0$, $\mathbf{x}_i^1$, $\mathbf{x}_i^2$,... $\mathbf{x}_i^T$ where T is the final time step. By gradually increasing the noise level in the points through the forward process, the final state of the point cloud $\mathbf{X}^T = \{ \mathbf{x}_i^T \}_{i=1}^{1084}$ eventually resembles samples from a standard normal distribution. 

Given the state of a point $\mathbf{x}_i^{t-1}$ at time step $t-1$, the probability distribution $q(\mathbf{x}^{t}_i | \mathbf{x}^{t-1}_i)$ for the next state $\mathbf{x}_i^t$, after addition of noise at time step $t$, is described by a Gaussian kernel:
\begin{equation}
    q(\mathbf{x}^{t}_i | \mathbf{x}^{t-1}_i)=\mathcal{N}\left(\mathbf{x}^{t}_i \big| \sqrt{1 - \beta_t} \mathbf{x}^{t-1}_i, \beta_t \mathbf{I}\right)
\end{equation}
with a mean of $\sqrt{1 - \beta_t} \mathbf{x}^{t-1}_i$ and variance of $\beta_t \mathbf{I} $. Here $\beta_t$ is a  hyperparameter controlling the amount of noise added at each time step and $\mathbf{I}$ is the identity matrix with dimensions matching the particle vector $\mathbf{x}_i$.

The reverse anti-diffusion process is the inverse of the forward process which transforms Gaussian noise samples to the final point cloud. Unlike the forward diffusion process, the reverse process requires training a neural network to identify and remove the noise added to the state in each, corresponding forward step. Starting with a sample from the standard normal noise distribution $\tilde{q}(\mathbf{x}^{T}_i) $=$\mathcal{N}(0,I)$, which approximates $q(\mathbf{x}^{T}_i)$, a neural network sequentially predicts the de-noised state in the previous time step, finally recovering the original point $\mathbf{x}_i^0$ in the point cloud.

Given the state $\mathbf{x}^{t}_i$ and the condition vector $z$, the probability for the previous state $\mathbf{x}^{t-1}_i$ is also given by a Gaussian kernel: 
\begin{equation}
\tilde{q}_{\theta}(\mathbf{x}^{t-1}_i | \mathbf{x}^{t}_i, z) = \mathcal{N}(\mathbf{x}^{t-1}_i | \mu_\theta(\mathbf{x}^{t}_i  , t, z), \beta_t I),
\label{eqtheta}
\end{equation}

where $\mu_{\theta}$ is the de-noised mean as predicted by a neural network with parameters $\theta$, based on the current state $\mathbf{x}^{t}_i$, the time step $t$, and the latent vector $z$.

\subsection{Training and Generation}

The PointNet-based encoder, the normalizing flow-based decoder and the probabilistic diffusion model are trained together to minimize the objective function given by: 

\begin{align}
    \mathcal{L} = 
    \mathbb{E}_q \bigg[ &
    \sum_{t=2}^{T} \sum_{i=1}^{N} 
    D_{KL} \Big(q(\mathbf{x}_{i}^{t-1} | \mathbf{x}_{i}^{t}, \mathbf{x}_{i}^{0}) \Big\| 
    \tilde{q}_{\theta} (\mathbf{x}_{i}^{t-1} | \mathbf{x}_{i}^{t}, z) \Big) \notag \\
    & - \sum_{i=1}^{N}  \ln \tilde{q}_{\theta} \Big(\mathbf{x}_{i}^{0} | \mathbf{x}_{i}^{1}, z \Big) + D_{KL} \Big( \tilde{q}_{\phi}(z | \mathbf{X}^{0}) \Big\| p(z) \Big)
    \bigg].
   \label{eqobj}
\end{align}

Here, $\mathbb{E}_q$ stands for the expectation value with respect to the distribution of $q$ and $D_{KL}$ stands for the Kullback-Leibler divergence, which quantifies the difference between the two distributions.

The first term measures the KL divergence between the true posterior for state $q(\mathbf{x}_{i}^{t-1})$, given previous state $\mathbf{x}_{i}^{t}$, and the undisturbed UrQMD data $\mathbf{x}_{i}^{0}$ and the posterior approximated by the diffusion model $\tilde{q}_{\theta}(\mathbf{x}_{i}^{t-1})$, given the state $\mathbf{x}_{i}^{t}$ and condition vector $z$. As discussed in \cite{ho2020denoising}, the true posterior for a point can be computed to be a Gaussian whose mean and variance depends only on the undisturbed state $\mathbf{x}_{i}^0$, current state $\mathbf{x}_{i}^{t}$  and the hyperparameter $\beta_t$ that controls the noise added in each step. 

Equation \ref{eqtheta} gives the approximate posterior $\tilde{q}_{\theta} (\mathbf{x}_{i}^{t-1} | \mathbf{x}_{i}^{t}, z)$. The summation iterates over all time steps, from $t=2$ to $T$, and across all points, $i=1$ to $N$, computing the KL divergence for the entire point cloud across all time steps.

The second term calculates the negative log-likelihood of the original UrQMD point cloud, given the noisy point cloud at time step $t=1$ and the condition vector $z$, which is given also by equation \ref{eqtheta}.

The last term in the objective function calculates the KL divergence between the posterior for condition $z$, given the UrQMD event $\mathbf{X}^{0}$ and is given by equation \ref{eqphi}. Finally, the probability distribution of $z$ is given by equation \ref{eqz}. 

For more details on the loss function and the training algorithm, we refer to the original implementation of probabilistic diffusion model \cite{ho2020denoising}  and point cloud diffusion model \cite{Luo_2021_CVPR}.

After successful training, only the normalizing flow-based decoder and the diffusion models are necessary for generation of point clouds. This part of the network is highlighted with gray background in figure \ref{heidi}. The PointNet encoder is necessary only when training the model.  

\section{Results}
\emph{HEIDi} is trained to minimize the objective function defined in equation \ref{eqobj}. After training, the model is evaluated for its capability to generate "realistic collision events" as described by UrQMD. To assess this, various distributions of distinct particle species from 50000 \emph{HEIDi}-generated events are compared with 50000 UrQMD-generated events.

\emph{HEIDi} generates a list of 1084 particle vectors per event, where each vector comprises of the 3 dimensional momentum vector and the unique ID which identifies the respective hadron species. Removing all the fake particles (ID= 0) in a generated event gives the final state output particles in that event. Although \emph{HEIDi} generates particles from all 26 different hadron types, for these comparisons we focus on the 16 most abundant hadron species. The remaining 10 hadron types have a too low event multiplicities (i.e., mean multiplicities $\lessapprox 10^{-2}/$event).

\begin{figure*}[]
   \includegraphics[width=\textwidth]{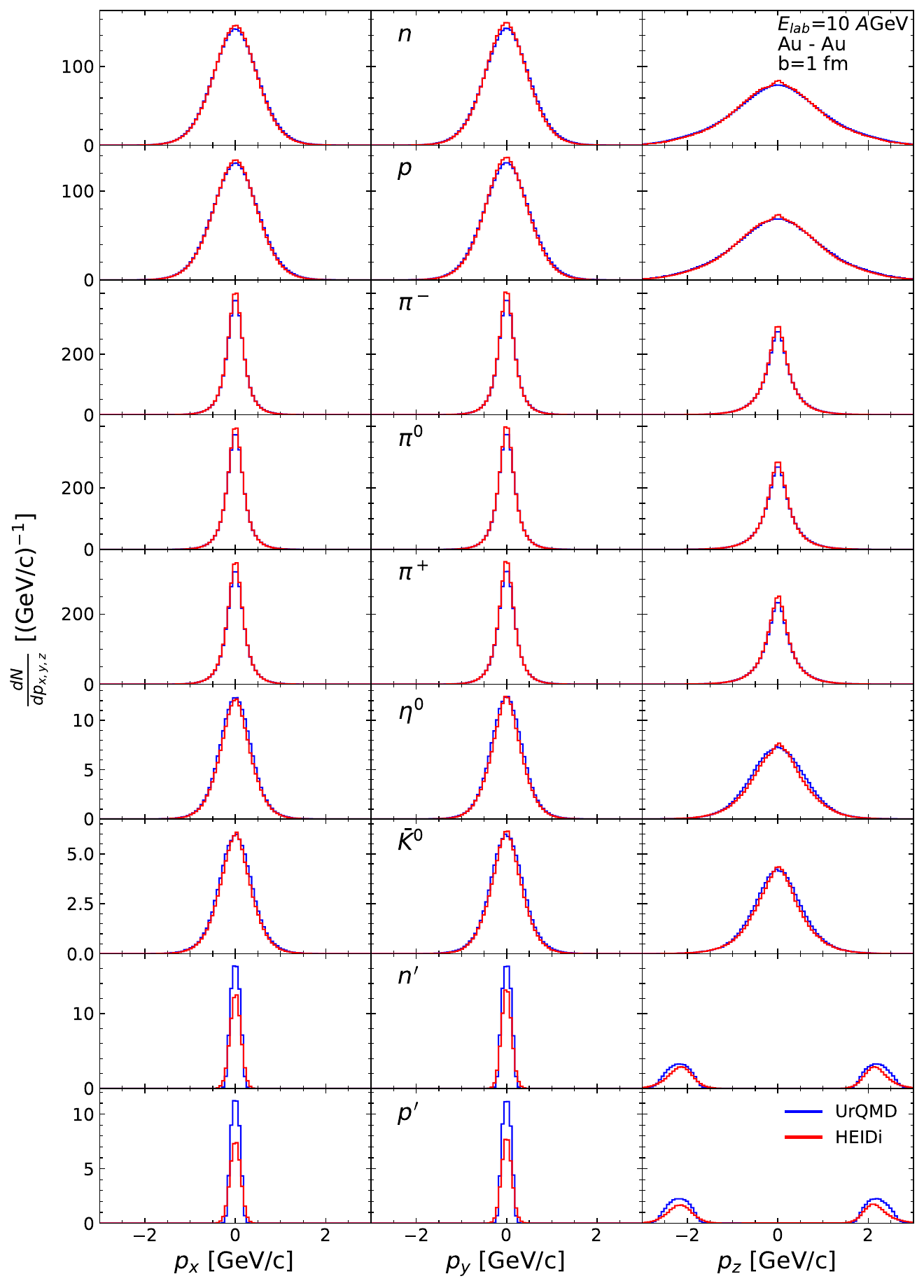}
   \caption{(Color online) Momentum distributions of various hadron types generated by \emph{HEIDi} for Au-Au collisions with $b=1$ fm at 10 $A$GeV. Each row corresponds to one hadron type. The columns 1, 2 and 3 represent the distributions for $p_x$, $p_y$, and $p_z$ respectively. UrQMD and  \emph{HEIDi} results are shown in blue and red curves respectively. }
   \label{mom1}
 \end{figure*}

\begin{figure*}[]
   \includegraphics[width=\textwidth]{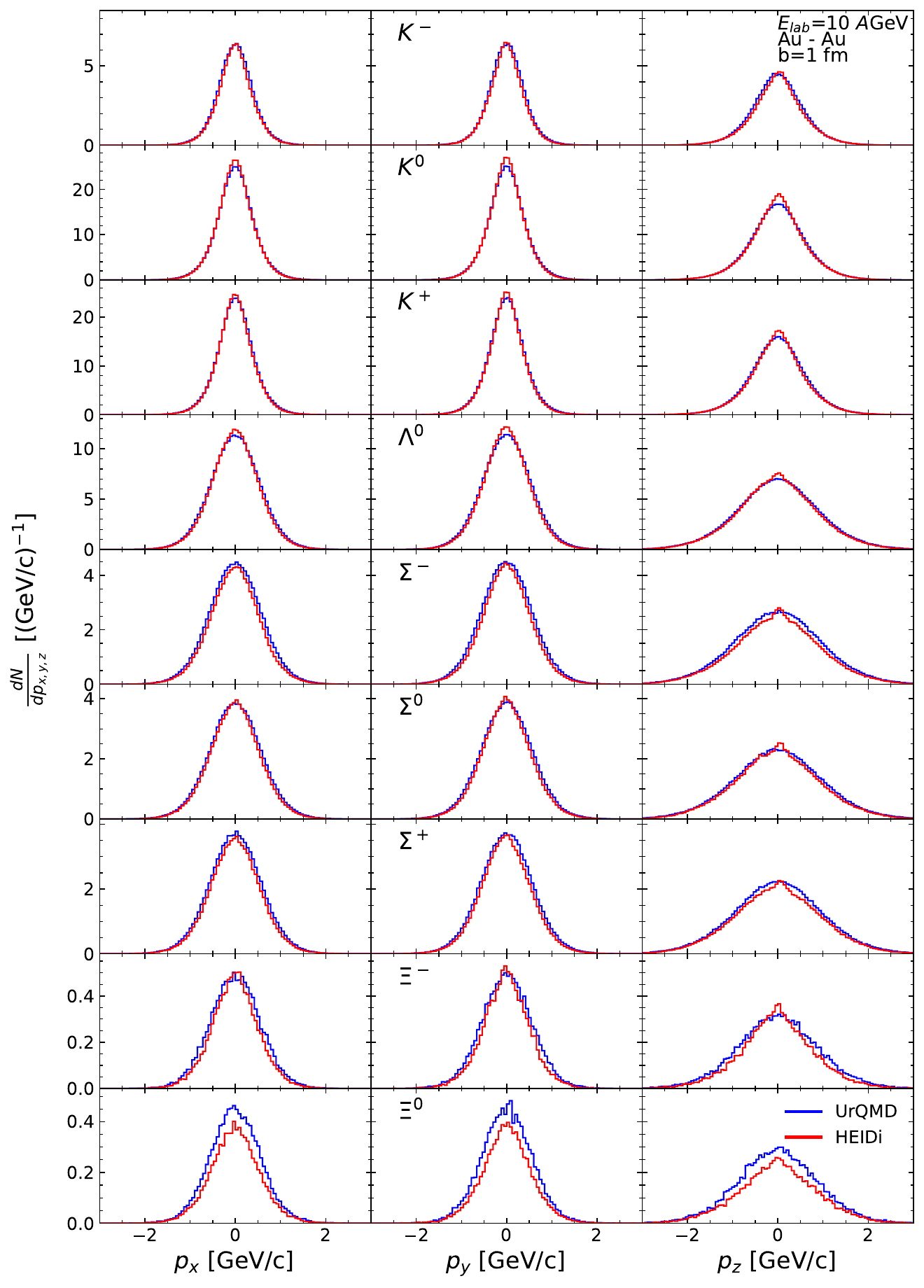}
   \caption{(Color online) Momentum distributions of various hadron types generated by \emph{HEIDi} for $b=1$ fm, Au-Au collisions at 10 $A$GeV. The color scheme and plot layouts are similar to figure \ref{mom1}.}
   \label{mom2}
 \end{figure*}

A comparison of the $p_x, p_y$ and $p_z$ distributions for various particle types generated by \emph{HEIDi} and UrQMD are shown in figures \ref{mom1} and \ref{mom2}. Evidently, the momentum distributions of various hadron types for particles generated by \emph{HEIDi} are in excellent agreement with those by UrQMD simulations. Despite slightly underestimating very small values of $p_x$ and $p_y$, of spectator nucleons, \emph{HEIDi} has successfully learned the $p_z$ distributions of spectators, which differ drastically from the distributions of participant nucleons and from all other hadron species.

\begin{figure*}[]
   \includegraphics[width=\textwidth]{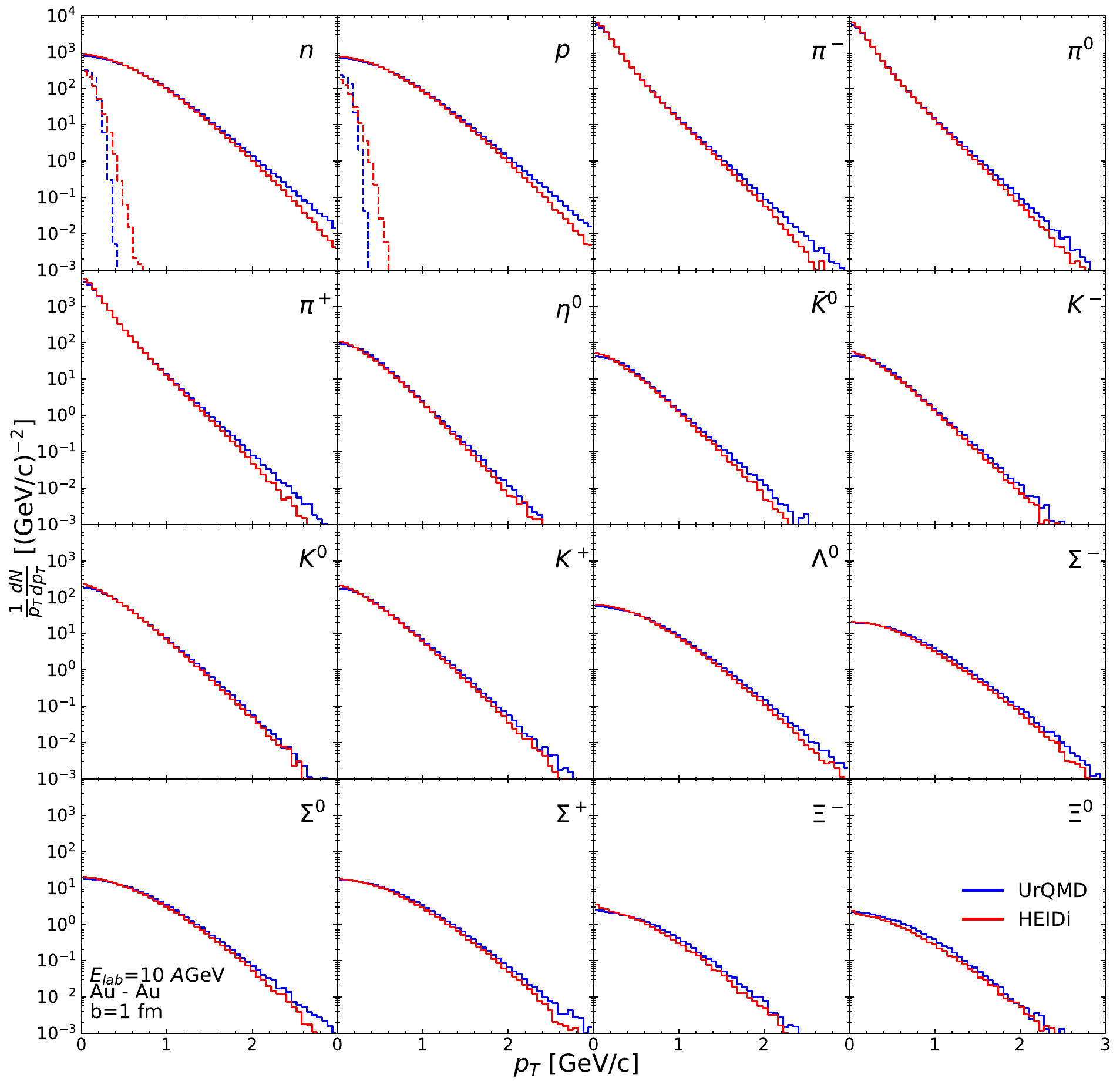}
   \caption{(Color online) Transverse momentum distribution of various hadrons. The results are for 10 $A$GeV Au-Au collisions with impact parameter $b=1$ fm. The \emph{HEIDi} distributions are shown as red curves while the UrQMD distributions are shown in blue. The dashed curves in the upper left two plots are the $p_T$ distributions for the spectator nucleons where the solid curves in the plots show the $p_T$ distributions for participant nucleons.}
   \label{ptfig}
 \end{figure*}

Apparently, \emph{HEIDi} performs well in learning the individual components of the momenta of various hadron species. It is interesting to see how well the joint probabilities of the momentum components are learned. Figure \ref{ptfig} shows the the transverse momentum distributions of various hadrons. \emph{HEIDi} learns accurately all $p_T$ distributions for different hadrons. With only a moderate training dataset, \emph{HEIDi} efficiently captures the complex joint probability distributions, across \emph{six orders in magnitude} for all hadron species. 

\begin{figure*}[]
   \includegraphics[width=\textwidth]{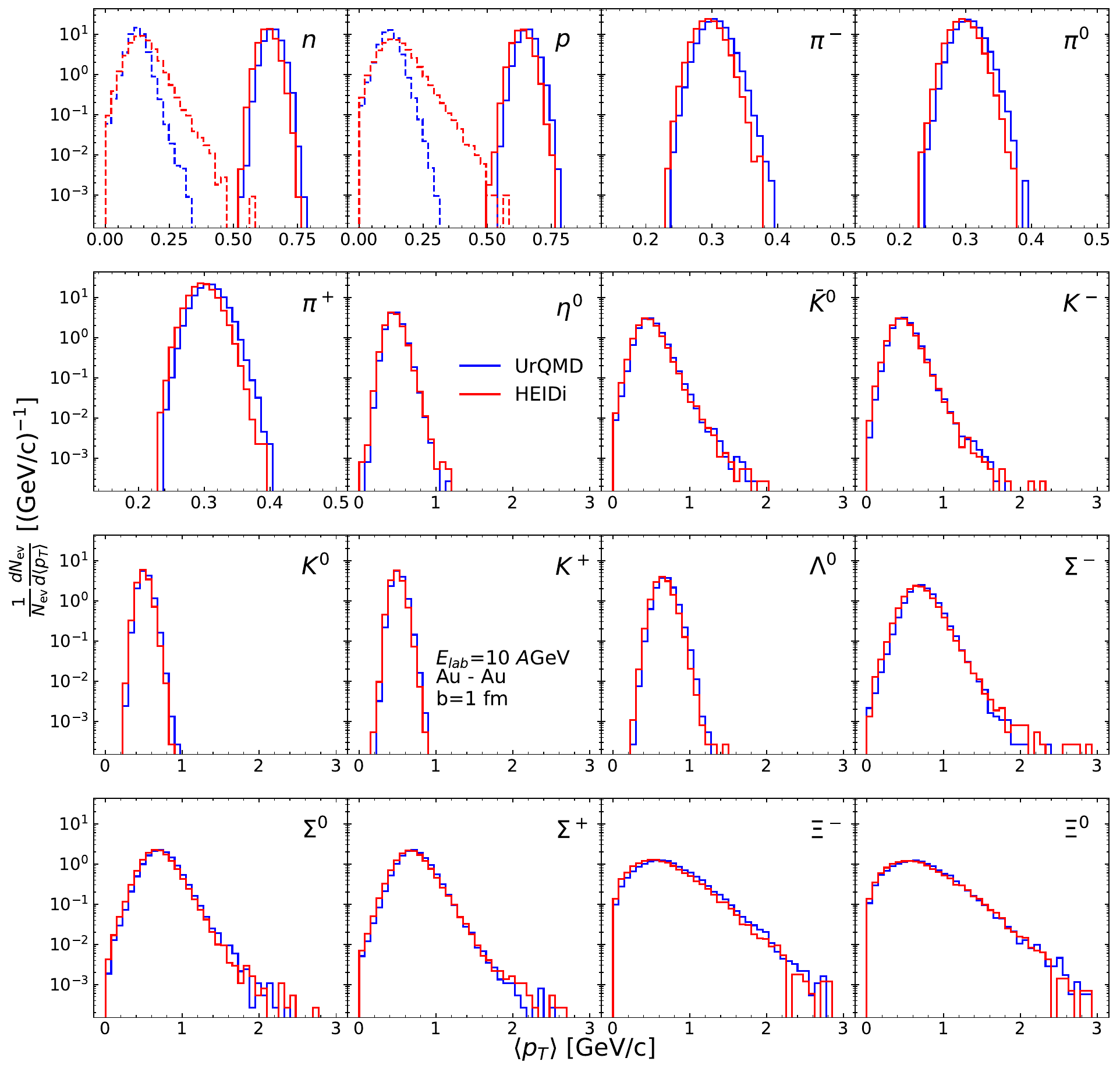}
   \caption{(Color online) $\langle p_T \rangle$ distribution of various hadrons for Au-Au collisions with $b=1$ fm at 10 $A$GeV. The \emph{HEIDi} distributions are shown as red curves, while the UrQMD distributions are shown in blue. The dashed curves in the upper left two plots are the distributions for spectator nucleons and the solid curves in the plots represent the distributions for participant nucleons.}
   \label{mptfig}
 \end{figure*}

In addition to learning various features of individual particles generated in each event, it is very important that \emph{HEIDi} captures various global properties of events such as the specific mean transverse momenta $\langle p_T \rangle$ and the average multiplicities of each hadron species separately in each and every event. The $\langle p_T \rangle$ distributions of various hadrons are presented in figure \ref{mptfig}. 

A good agreement between the $\langle p_T \rangle$ distributions of \emph{HEIDi} events and UrQMD events for all different hadron species can be noted. For certain particle species and at very high $\langle p_T \rangle$, where the training data samples are rather limited, \emph{HEIDi} learns the distribution so well that only moderate deviations from UrQMD distributions are observed.

\begin{figure*}[]
   \includegraphics[width=\textwidth]{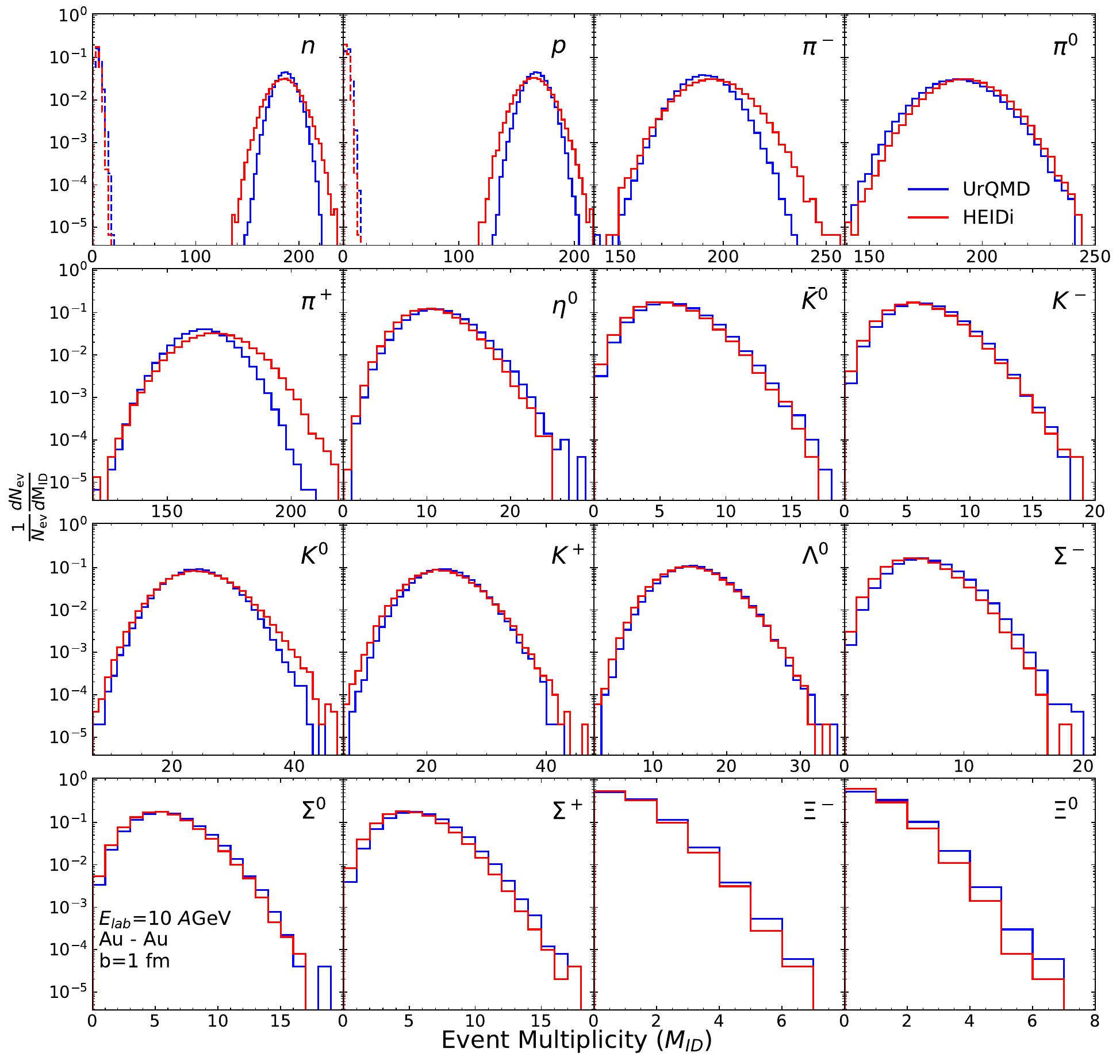}
   \caption{(Color online) Multiplicity distribution of various hadron species for Au-Au collisions with $b=1$ fm at 10 $A$GeV. In the upper left two plots, the distributions for spectator nucleons are shown as dashed curves while the solid curves in the plots denote the distributions for participant nucleons. The multiplicity distribution constructed from \emph{HEIDi} events are shown as red curves while the distributions from UrQMD events are shown in blue. }
   \label{multi}
 \end{figure*}

Figure \ref{multi} presents the multiplicity distributions of different hadron species. \emph{HEIDi} distinguishes the differences in the multiplicities of various hadron types in each event, accurately learning the relative probabilities. \emph{HEIDi} also recognizes the large difference in the multiplicity distributions for spectators and participants, thus well modeling the event-by-event fluctuations of the respective interaction volumes. For certain hadron types, \emph{HEIDi} slightly overestimates the probability for very high and very low multiplicities. 

Reproducing the multiplicity distributions are much more difficult than reproducing the rapidity distributions or momentum distributions, as it requires accurately capturing small, many-particle correlations in individual events. The event-by-event multiplicity of the particles which show these deviations such as neutrons, protons, and pions also have significantly higher variance compared to those of rarer particles.  Learning high-variance distributions of more abundant particles from a limited dataset is more challenging as here the many body correlations are more important statistically. If a particle is very rare, then its multiplicity distribution will approach the poissonian limit which is easy for a model to learn. For particles that are very abundant the contribution from many-particle correlations becomes more and more prevalent and thus is harder to capture. Although our model is designed to capture such correlations, the diffusion model is informed of these correlations by the condition vector generated by the normalizing flow decoder rather than hard constraints. Thus, it is easier for the model to learn the less correlated low variance distributions of the rare particles. We also find that increasing the diversity of training data by conditioning the model on different collision centralities significantly improves the overall multiplicity distributions of these abundant particles for various centralities. This is discussed in section \ref{cond}.

It is important to emphasize that, generative models are usually not benchmarked on the multiplicity distributions but rather on either being able to create single instances or event-averaged distributions. In our work, this has already been demonstrated in Figure \ref{mptfig}, where we observe excellent agreement with UrQMD over more than five orders of magnitude in probability for all considered particle species.

\begin{figure*}[]
   \includegraphics[width=\textwidth]{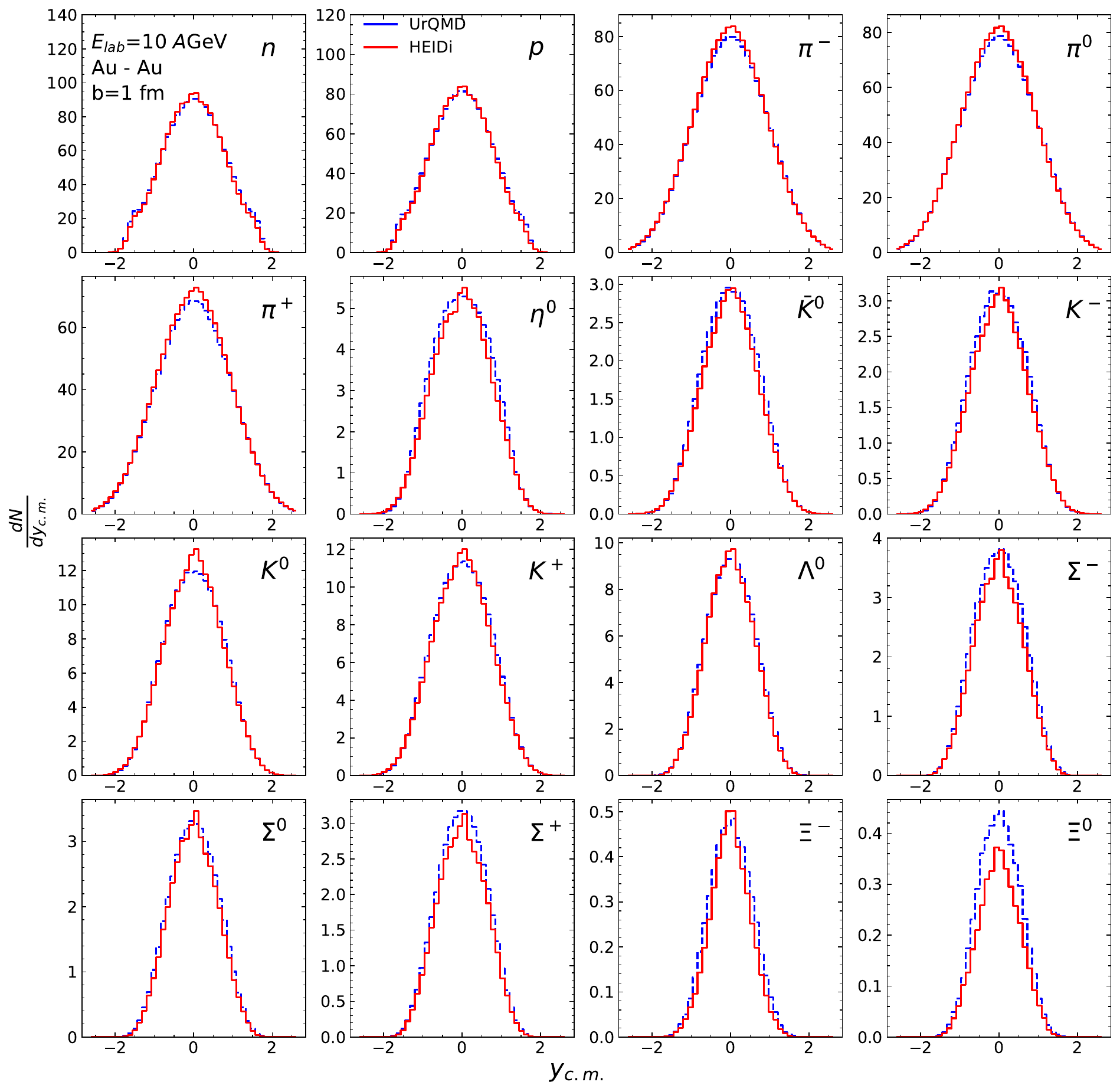}
   \caption{(Color online) Rapidity distribution of various hadron species for Au-Au collisions with $b=1$ fm at 10 $A$GeV. The distribution constructed from \emph{HEIDi} events are shown as red curves while the distributions from UrQMD events are shown in blue. The upper left two plots show the rapidity distributions for all neutrons and all protons (including spectators) respectively. }
   \label{rapi}
 \end{figure*}

 The rapidity distributions of various hadrons generated by \emph{HEIDi} are compared with those from UrQMD simulations in figure \ref{rapi}. The particles generated by \emph{HEIDi} follow  the rapidity distributions of hadrons in UrQMD data well. For some hadrons, a slight but visible excesses of particles are found in \emph{HEIDi} events for small rapidities and $p_z$ close to 0. \emph{HEIDi} tends to overestimate the number of very low momentum particles as their number in the current training dataset is quite limited. However, this only leads to a small effect on the total yield.
  
\begin{figure*}[]
   \includegraphics[width=\textwidth]{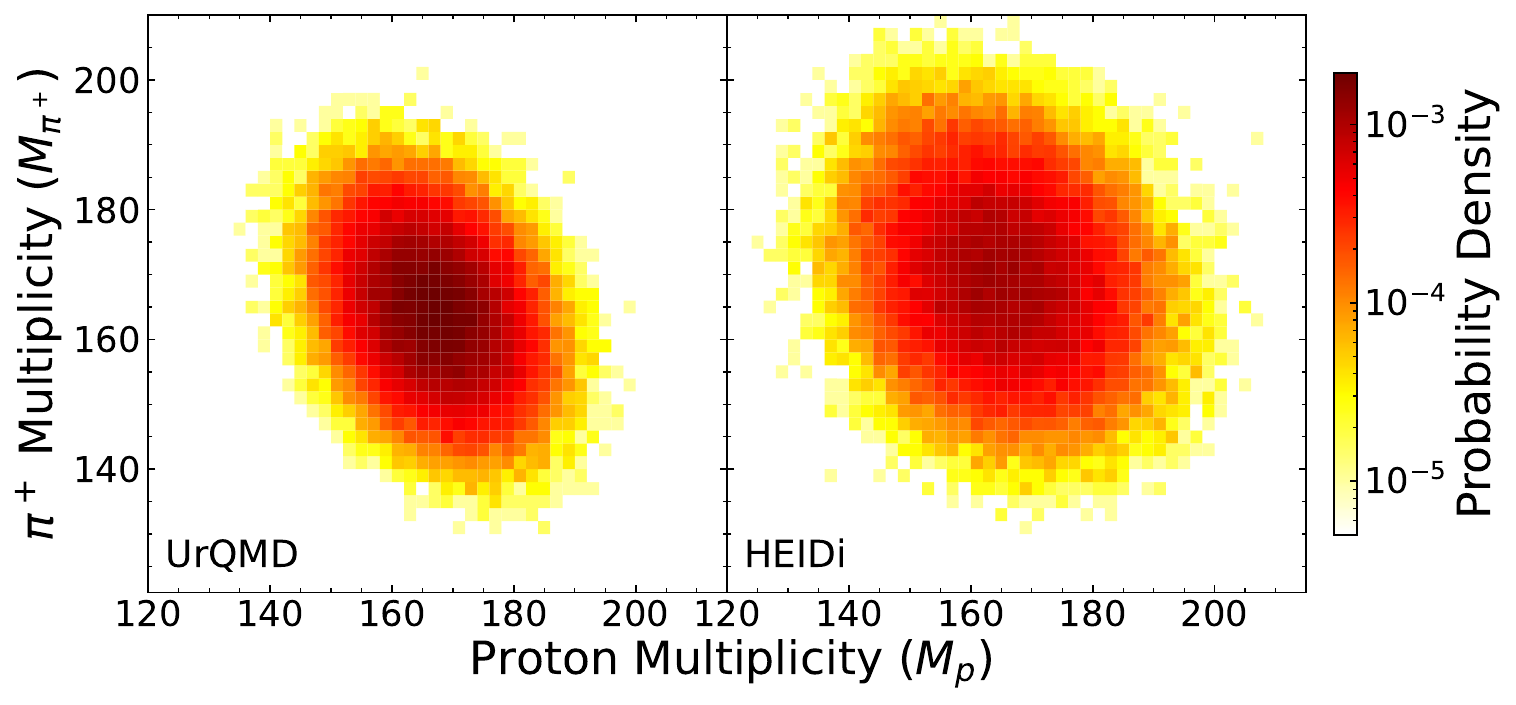}
   \caption{(Color online) Event-by-event multiplicity correlation between protons and $\pi^+$ in Au-Au collisions for $b = 1$~fm at 10~$A$GeV. The left panel shows the true correlation as obtained from UrQMD events, while the right panel shows the corresponding correlation learned by \emph{HEIDi}. }
   \label{multcorr}
 \end{figure*}

We have already demonstrated that the model accurately reproduces the marginal distributions of various features, as well as various momentum space correlations, as reflected in the rapidity and transverse momentum distributions. To further assess the model's ability to capture event-level correlations, we investigate the event-by-event multiplicity correlation between protons and $\pi^+$, as shown in Figure~\ref{multcorr}. The model successfully captures the overall negative correlation between the proton and $\pi^+$ multiplicities in an event. While the learned correlation does not perfectly match the true correlation from UrQMD, this is expected, as accurately learning detailed event-by-event correlations remains a challenge for generative models. These deviations in learned correlations, in turn, contribute to the discrepancies observed in the multiplicity distributions presented in Figure~\ref{multi}.

 Despite being trained on a relatively small dataset of 18000 events, \emph{HEIDi} successfully captured most of the key properties and correlations among distinct hadron types, as well as global event characteristics. With this limited dataset, \emph{HEIDi} is able to generate "UrQMD like events" with the components of momentum vector or the transverse momentum of generated particles closely following the UrQMD distributions even in the tails where the training samples are sparse. The small dataset was chosen to demonstrate \emph{HEIDi}'s ability as an AI emulator to learn underlying probability distributions and correlations from limited samples to ultimately generate large number of realistic samples quickly to achieve a higher statistical significance. However, capturing all correlations in UrQMD, particularly very rare or absent ones in the training data, is inherently difficult. This limitation is reflected in discrepancies in the tails of the multiplicity distribution and in the number of very low-momentum particles, indicating regions of phase space where the model might be undertrained. A better training strategy or a diverse training set would be necessary to improve the accuracy of the model in these regions.

The primary motivation to develop a neural network-based event-by-event emulator for the known traditional heavy-ion collision simulation models is to overcome the computational bottleneck due to the large computation times which prevent us from performing necessary, expensive parameter estimation tasks. When deployed on an Nvidia A100 GPU with 40 GB of memory, \emph{HEIDi} generates an event in about 30 milliseconds, whereas the UrQMD cascade model takes approximately 3 seconds per event. This speedup of about 2 orders of magnitude is truly remarkable as it is straightforward to train \emph{HEIDi} on similar input data from more complex and expensive 3+1-dimensional dynamical models, such as UrQMD with potentials, complex in-medium momentum dependent cross sections etc. or the hybrid model (including hydrodynamic phase) of UrQMD. Here speedups of at least five orders of magnitude are expected. \emph{HEIDi}, trained on such datasets, will be fast enough for a comprehensive Bayesian inference of the EoS or any other physics parameters, for which numerous, multi-differential observables will necessarily be calculated. \emph{HEIDi} is fully differentiable: the gradients of the objective function with respect to every single model parameter can be calculated. Hence, novel optimization techniques can be explored to extract the underlying model parameters directly from event-by-event experimental data, without the use of pre-defined observables.

\section{Conditioning the Model on Collision Properties}\label{cond}

For the model to be practically useful in tasks such as Bayesian inference, the model should be able to generate collision events with varying properties such as beam energy, centrality, and the equation of state. We have already demonstrated that the model accurately generates collision event outputs for a fixed centrality ($b=1$ fm) Au-Au collisions at 10 $A$GeV. Extending the model to handle events with different collision features is straightforward and can be achieved by extending the latent vector $z$ with necessary collision properties. The required condition (collision property) can be added as an additional non-trainable parameter  to the latent vector $z$ generated by the normalizing flow decoder.

\begin{figure*}[]
\includegraphics[width=\textwidth]{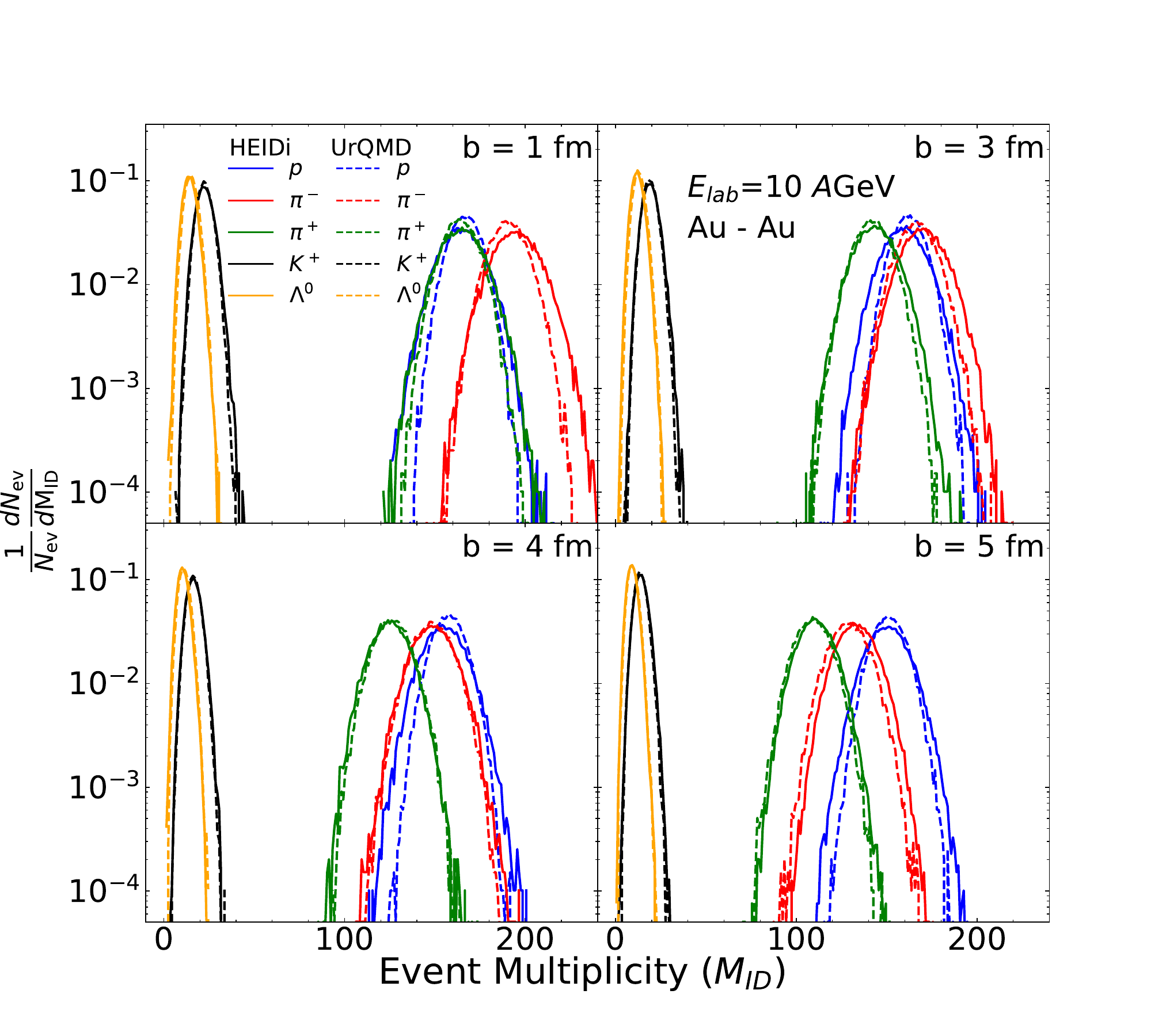}
\caption{(Color online) Multiplicity distributions of selected hadron species for Au-Au collisions at 10 $A$GeV for impact parameters $b = 1$ fm (top left), 3 fm (top right), 4 fm (bottom left), and 5 fm (bottom right). Distributions generated from \emph{HEIDi} events are shown as solid curves while the distributions generated from UrQMD events are shown as dashed curves.}
\label{multib}
\end{figure*}

To illustrate this capability, we trained the \emph{HEIDi} architecture on 30000 Au-Au events at 10 $A$GeV, comprising 10000 events each for impact parameters of 1, 3, and 5 fm. The impact parameter was included as a conditioning variable by extending the latent vector $z$ accordingly. The model was then tested on UrQMD events with impact parameters of 1, 3, 4, and 5 fm. The resulting multiplicity distributions for selected hadron species are shown in Figure~\ref{multib}.

It can be seen that the model successfully captures the centrality dependence of hadron multiplicities. While some discrepancies remain at $b = 1$ fm, there is a good agreement with UrQMD results for the other tested impact parameters. Furthermore, the model also demonstrates strong interpolation capabilities, as evident in the accurate reproduction of multiplicities at $b = 4$ fm, a value not included in the training set. The corresponding rapidity distributions for selected hadron species across various impact parameters are presented in \cite{OmanaKuttan:2024mwr} and show an excellent agreement between \emph{HEIDi} and UrQMD for all tested centralities. These results indicate that training the model on a more diverse and extensive dataset enhances the model's generalization ability and predictive power. Furthermore, they demonstrate the flexibility and adaptability of the proposed model in handling additional conditions such as various collision parameters without any modification to the core architecture.

\section{Applications in heavy-ion physics and beyond}
\emph{HEIDi} as a robust framework for comprehensive Bayesian inference or other parameter estimation tasks can be deployed for various important applications within high-energy heavy-ion collisions and other detector-intensive experiments. \emph{HEIDi} uses a point cloud data structure, which is extremely flexible and easily adaptable for diverse tasks in high-energy physics. Developers of different models or detector simulations can readily adapt \emph{HEIDi} to build their own generative AI simulation frameworks trained on specific parameter sets or physical features, thereby enabling rapid generation of large-scale event samples without requiring detailed knowledge of the underlying code of the physics model.  In addition, experimental collaborations can deploy models based on \emph{HEIDi} for fast, online data comparison, which can be useful for first-level event analysis, triggers for interesting physics, calibration, quality control, and real-time monitoring, fault or anomaly detection in detector systems.

Conditional point cloud diffusion has a wide range of potential applications beyond heavy-ion physics. Architecture similar to \emph{HEIDi} can easily perform analyses and generative tasks for scientific data processing wherever data are collected electronically or with multiple sensors with point cloud structure. \emph{HEIDi} demonstrates the feasibility of conditional generation for high-dimensional, categorical point clouds. Similar models can find important applications across various other fields which directly benefit mankind, such as biomedical imaging and diagnosis, predictive maintenance for technological hardware, early warning systems for natural hazards, environmental surveillance, urban planning \cite{Li:2024fiw}, and autonomous systems.

\section{Conclusions}
This work introduces \emph{HEIDi}, a conditional point cloud generative model for event-by-event collision output in relativistic heavy-ion collisions. Based on \cite{Luo_2021_CVPR}, \emph{HEIDi} uses a PointNet-based point cloud encoder and a normalizing flow-based decoder to create a latent encoding which is used as a condition for a probabilistic diffusion model that generates complete collision output. When trained on UrQMD cascade data, the model learns to generate "realistic" point clouds of collision output which contains 26 most abundant distinct hadron species. Particles generated by the DL model accurately reproduce different probability distributions of UrQMD data. The \emph{HEIDi} model not only capture various correlations among different particles in each event but also learns well the different event level properties. The generated hadrons are shown to successfully reproduce the UrQMD distributions, both for different components of  momentum vectors $p_x$, $p_y$,  $p_z$, transverse momenta $p_T$, and multiplicities $M$ as well as  rapidities $y_{cm}$. \emph{HEIDi} generates events two order of magnitude faster than the conventional UrQMD cascade computations.

Capturing all correlations within an event in high-energy collisions remains a significant challenge for current point cloud-based generative models, including \emph{HEIDi}. This limitation is particularly evident in the deviations observed at the tails of the multiplicity distributions for nucleons and pions. Simply increasing the size of the training dataset does not directly lead to substantial improvements in the multiplicity distributions. However,it is found that training \emph{HEIDi} on a more diverse and extended dataset by including multiple centrality  events improves overall agreement with UrQMD for the event-by-event multiplicity distributions of nucleons and pions for various centralities. The same strategy can be easily extended to incorporate a wide range of physical conditions, including collision energy, system and asymmetry, and parameters governing the equation of state (EoS), without requiring any changes to the underlying model architecture. In addition, we are currently exploring other techniques to enhance the model's ability to capture correlations from limited training data. These include  multi-step generation approaches such as first generating particle multiplicities and then sampling the particle momenta and conditioning the model directly on initial-state profiles (e.g., nucleon positions and momenta).

The architecture of \emph{HEIDi} can be adapted for any other event-by-event collision output generation as possibly dictated by any other physical model. The choice of the point cloud format to represent the event-by-event collision data makes it flexible for accelerating detector simulations as well. Such extensions of \emph{HEIDi} will be tackled in future research.  

\emph{HEIDi}'s high statistics heavy-ion event generation marks a significant milestone in AI-accelerated event-by-event heavy-ion simulations. Through ultra-fast generation of complete heavy-ion events in point cloud format, \emph{HEIDi}-based models will allow us to use computationally expensive observables for complex parameter estimation tasks, as well as enable ultra-fast large-scale simulations for various theoretical studies and experimental applications of interest. As a robust and flexible generative AI framework, \emph{HEIDi} lays the ground for a first foundation model for heavy-ion collisions. 

\textbf{Acknowledgment.} This work is supported by the the BMBF under the KISS project (M.O.K, K.Z), the SAMSON AG (J.S, K.Z), the CUHK-Shenzhen university development fund under grant No. UDF01003041 and UDF03003041 (K.Z), and Shenzhen Peacock fund under No. 2023TC0179 (K.Z) and the Walter Greiner Gesellschaft zur F\"orderung der physikalischen Grundlagenforschung e.V. through the Judah M. Eisenberg Laureatus Chair at the Goethe Universit\"at Frankfurt am Main (H.S).

\bibliography{bib_new}
\bibliographystyle{apsrev4-1}

\end{document}